\documentstyle[11pt,a4,epsfig,amssymb]{article}

\makeatletter
\@addtoreset{equation}{section}
\makeatother

\newcommand{\comment}[1]{{}}

\newcommand{\beq}{\begin{equation}}
\newcommand{\eeq}{\end{equation}}
\newcommand{\ba}[1]{\begin{array}{#1}}
\newcommand{\ea}{\end{array}}
\newcommand{\bea}{\begin{eqnarray}}
\newcommand{\eea}{\end{eqnarray}}

\title{
\begin{flushright} {\normalsize \today} \end{flushright}
\vspace{2cm}
{\bf
Higgs Mass Determination from Direct Reconstruction 
at a Linear $e^+e^-$ Collider
} \\
\vspace{1ex}
}

\date{}

\author{
\large {\sc Aurelio Juste\/} \\
\\
\normalsize  Fermi National Accelerator Laboratory, \\
\normalsize  P.O. Box 500, MS 357, \\
\normalsize  Batavia, IL 60510, \\
\normalsize  Phone: 1\,(630)\,840\,-6565 \ \ Fax: 1\,(630)\,840\,-8481 \\
\normalsize  e-mail: juste@fnal.gov \\
\\
}

\begin{document}

\maketitle
%\vspace{1ex}
\begin{abstract}
%\vspace{2ex}
%\begin{center} PACS'96: 02.50.Ph, 07.05.Kf, 07.05.Mh \end{center}
%\vspace{1ex}
We study the feasibility of a precise measurement
of the mass of a 120 GeV MSM Higgs boson
through direct reconstruction of $ZH \to q\bar{q}H$ events
that would be achieved in a future $e^+e^-$ linear collider 
operating at a center-of-mass energy of 500 GeV.
Much effort has been 
put in a ``realistic simulation'' by including irreducible and reducible
backgrounds, realistic detector effects and reconstruction procedures
and sophisticated analysis tools involving Neural Networks and
kinematical fitting. As a result, the Higgs mass is determined with 
a statistical accuracy of 50 MeV and the Z-Higgs Yukawa coupling
measured to 0.7\%, assuming 500 fb$^{-1}$ of integrated luminosity.
\vspace{1ex}
\begin{center}
Results presented at the International Workshop on Linear Colliders (LCWS99)\\
Sitges, Barcelona, Spain, 28 April - 5 May 1999
\end{center}
\end{abstract}

\newpage

\section{Introduction}
The Standard Model~\cite{SM} (MSM) of electroweak and hadronic interactions 
has been successfully tested so far to an
extremely high degree of accuracy. However, one of its key elements, the Higgs mechanism,
remains to be tested experimentally.
It is through the interaction with the ground state Higgs field
that the fundamental particles acquire mass, which in turn sets the scale
of the coupling with the Higgs boson. Once the Higgs boson is found,
all its properties have to be accurately measured. In the MSM, the Higgs mass
is not predicted by the theory but the profile of the Higgs particle: decay width,
branching ratios and production cross-section, are uniquely determined once
$M_H$ is fixed, hence the importance of performing a precise measurement of the Higgs mass.
In order to establish experimentally that this particle has indeed the properties of
a Higgs boson, we need to prove that it is a scalar particle and that it arises from
a field with a vacuum expectation value which contributes to the $W$ and $Z$ masses~\cite{Peskin}.
The latter is achieved by determining the Higgs Yukawa couplings to the $Z$ and $W$ gauge bosons,
which in an $e^+e^-$ collider can be measured 
from the Higgstrahlung process: $Z^*\to ZH$~\cite{higgstrahlung} 
and the fusion processes: $W^*W^*,\: Z^*Z^* \to H$~\cite{fusion}. The scalar nature of this particle
can be verified from its production angular distribution.

In this work we have assumed that the MSM Higgs boson in the ``Intermediate Mass Region'' 
($M_Z \leq M_H \leq 2 M_W$) will have already been discovered at the present (LEP2, TeVatron)
or future (LHC, NLC) accelerators, and study the feasibility of a precise
measurement of its mass and Yukawa coupling to the $Z$ boson from the Higgstrahlung process.

This mass range is favored both experimentally: from global fits to electroweak
precision observables at LEP, SLC and TeVatron the upper limit $M_H < 260$ GeV at 95\% CL~\cite{lepewwg}
is derived, whereas from direct search at LEP2 $M_H > 95.2$ GeV at 95\% CL~\cite{limit};
and theoretically: the stability and triviality bounds constrain the MSM Higgs boson mass to
be in the range 130 GeV $\leq M_H \leq$ 180 GeV~\cite{djouadi}.
A Higgs boson in the ``Intermediate Mass Region'' would be relatively more
difficult to detect at LHC than a heavy Higgs: whereas a Higgs boson in the mass
range 150-700 GeV can be found straightforwardly in the decay $H\to ZZ\to 4\ell$,
the discovery of the light MSM Higgs boson would be based on $H\to \gamma\gamma$
and would require 100 fb$^{-1}$ of data (a year's running at the design luminosity).
Instead, a light Higgs boson might be discovered with less than 1 fb$^{-1}$
of integrated luminosity at a future $e^+e^-$ linear collider operating at
$\sqrt{s}=300$ GeV~\cite{previous}. However, the job of a linear $e^+e^-$ collider
would be rather to study in great detail the properties of the Higgs particle,
which can be uniquely be attained in the clean and very high luminosity 
($\int {\cal L}dt\geq 100$ fb$^{-1}$/year) environment expected.

As already mentioned, we will focus on the case of the MSM Higgs boson, which is equivalent,
for $M_H \leq 130$ GeV, to the case of the light $h$ MSSM Higgs boson close to the
decoupling regime (where the MSM and MSSM Higgs sectors look practically the same).
For the sake of definiteness, we will assume $M_H = 120$ GeV and concentrate
on the $Z$ hadronic decay mode: $e^+e^- \to ZH \to q\bar{q}H$. For $M_H = 120$ GeV, 
the Higgs decays dominantly to $b\bar{b}$ (BR($H\to b\bar{b})\simeq 72\%$), which
leads to multi-jet event topologies involving at least 2 $b$-jets in the final state.
Therefore, one of the crucial experimental aspects will be flavor tagging.
Most of previous studies have been
focussed on the leptonic Z decay mode since it is less affected by background and there is a
better intrinsic resolution on $M_H$ (through the recoil mass distribution). 
For instance, in~\cite{pablo} a statistical uncertainty of $(\Delta M_H)_{stat}\simeq 110$ MeV is obtained
from the combination of $ZH \to e^+e^-H,\:\mu^+\mu^-H$ channels at $\sqrt{s}=350$ GeV and assuming 
500 fb$^{-1}$ of integrated luminosity .The disadvantage is the
lack of statistics ($BR(Z\to \ell^+\ell^-) \simeq 10\%$) as compared to 
the hadronic decay channel ($BR(Z\to q\bar{q}) \simeq 70\%$). 
Therefore, it is fully justified to investigate to what extent it
is possible to make of the hadronic decay channel a competitive measurement.

\section{Experimental Strategy}
As already mentioned, the main production mechanisms of the MSM Higgs boson in $e^+e^-$ annihilation
are the Higgstrahlung process: $e^+e^-\to Z^* \to ZH$ (with a cross-section scaling as $1/s$ and
therefore dominating at low energy) and $WW$-fusion: $e^+e^-\to \nu\bar{\nu}W^*W^*\to \nu\bar{\nu}H$
(which dominates at high energies since the
cross-section scales as $\log(s/M_H^2)$ ). One order of
magnitude smaller than $WW$-fusion there is also the contribution from $ZZ$-fusion:
$e^+e^-\to e^+e^-Z^*Z^*\to e^+e^-H$. At $\sqrt{s}=500$ GeV and for 
100 GeV $\leq M_H \leq$ 200 GeV, the Higgstrahlung and $WW$-fusion have approximately
the same cross-section.

\begin{figure}[htbp]
\begin{center}
\mbox{
\epsfig{file=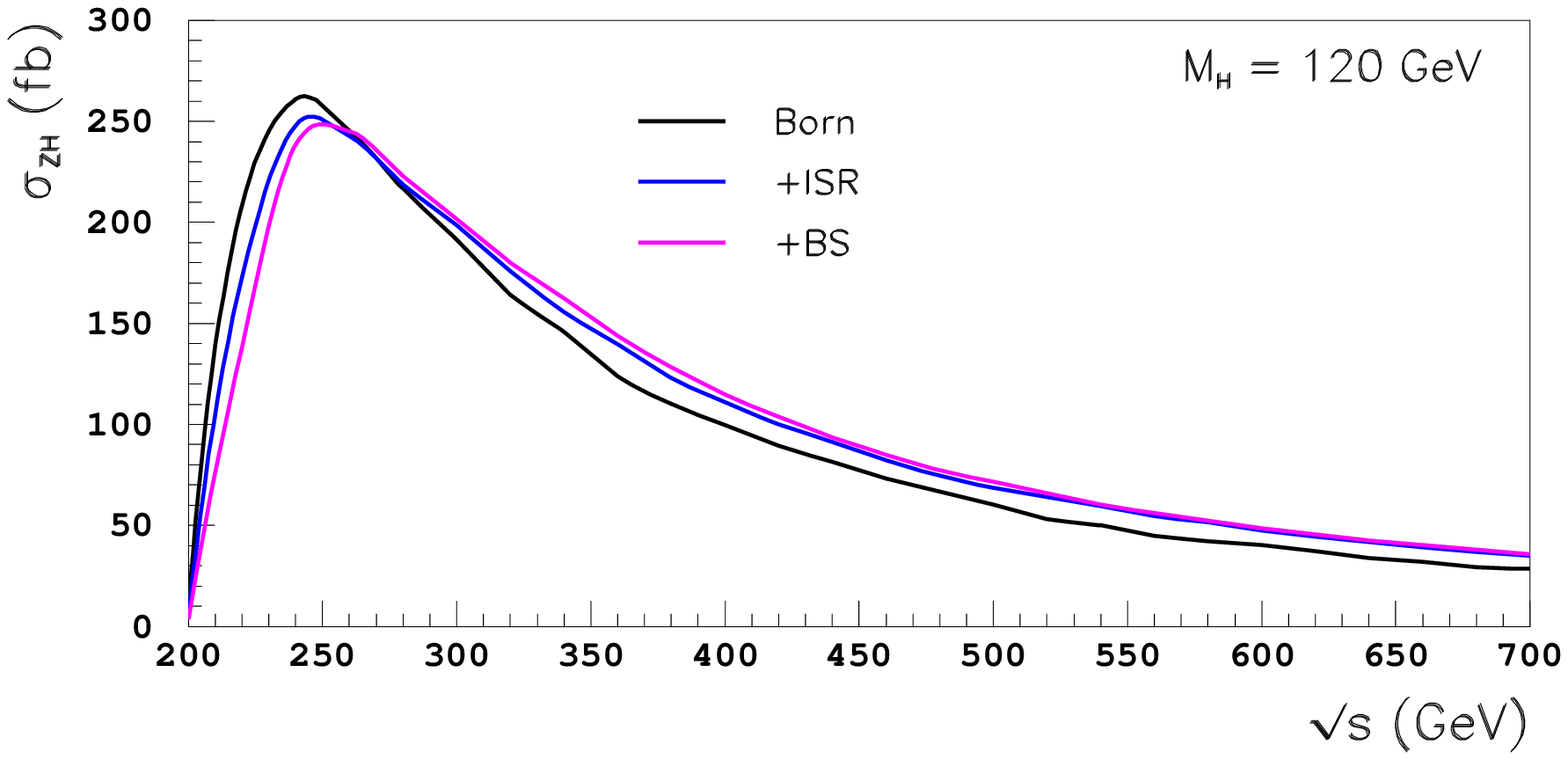,width=11cm}
}
\end{center}
\caption{\protect\footnotesize
Total cross-section for $ZH$ at lowest order, for $M_H = 120$ GeV,
as a function of the center-of-mass energy.
The effect of initial
state radiation and beamstrahlung on the total cross-section is also illustrated.}
\label{sigma1}
\end{figure}

The lowest order total cross-section for $ZH$, assuming $M_H = 120$ GeV,
is shown in Fig.~\ref{sigma1} as a function of the center-of-mass energy .
The effect of radiative processes in the initial state (initial
state radiation\footnote{Initial
state radiation will be used hereafter as synonymous with bremsstrahlung.} 
and beamstrahlung) on the total cross-section is also illustrated.
The total cross-section at $\sqrt{s}=500$ GeV is about 66 fb, which represents
an event sample of about 6600 events/year assuming\footnote{It has been assumed that 1 year's running = $10^7$ s.} ${\cal L}=10^{34}$ cm$^{-2}$s$^{-1}$.

The measurement is performed at $\sqrt{s}=500$ GeV. In principle, one would 
prefer to sit at the peak of $\sigma_{ZH}$:
\begin{eqnarray}
(\sqrt{s})_{peak} \simeq M_Z + \sqrt{2} M_H, \nonumber
\end{eqnarray}
\noindent that is $(\sqrt{s})_{peak} \simeq 260$ GeV for $M_H = 120$ GeV,
since the cross-section is enhanced by a factor $\simeq 3.4$ with respect
to $\sqrt{s}=500$ GeV and the background from $t\bar{t}$ is not present.
However, it is unclear how feasible it would be to collect enough integrated
luminosity at such a low center-of-mass energy in order to perform a
precise measurement.
Therefore, a more realistic strategy could be to perform
a first direct measurement of $M_H$ at the $t\bar{t}$ threshold ($\sqrt{s}=350$ GeV,
where still $\sigma_{ZH} (350 \:\rm GeV) \simeq 2.1 \; \sigma_{ZH} (500 \:\rm GeV)$) 
and contemporary with the top threshold measurements, and then perform the
precise measurement at $\sqrt{s} \geq 500$ GeV, where most of the integrated
luminosity will be collected.

There are two main strategies for the Higgs mass measurement:
\begin{itemize}
\item calculation of the mass recoiling against the $Z$: this has the nice feature of being
independent on assumptions about the Higgs decay modes~\cite{pablo}, and would show the
Higgs resonance even for invisible Higgs decays. 
The recoil mass is computed from
the reconstructed $Z$ 4-momentum assuming the nominal center-of-mass energy:
\begin{eqnarray}
M_H^2 = s-2\sqrt{s}E_Z+M_Z^2,
\end{eqnarray}
\noindent where $E_Z$ and $M_Z$ are, respectively, the $Z$ reconstructed energy and
invariant mass. Therefore, this method is more suitable for the $Z$ leptonic decay channel
since a more precise $Z$ reconstruction is possible.
\item direct reconstruction of the invariant mass of the Higgs decay products. As will
be shown, this method works better for hadronic $Z$ decays than the recoil mass method.
\end{itemize}

The total cross-section depends sensitively on the Z-Higgs Yukawa coupling, which can
thus be inferred from the comparison of the measured total cross-section with the
theoretical expectation as a function of $g_{ZZH}$. Therefore it is possible to verify whether
this field is the responsible for the whole $Z$ mass (as in the MSM), or only for a fraction of it.

\section{Simulation Aspects}
The signal and the different backgrounds have been generated with PYTHIA~\cite{jetset}. 
The top quark and Higgs masses have been assumed to be
$m_t = 175$ GeV and $M_H = 120$ GeV, respectively.
Interference between signal and background have been neglected.
The event samples have been generated
at $\sqrt{s} = 500$ GeV, including initial state radiation (ISR) and
beamstrahlung. For efficiency reasons, a generation cut $\sqrt{s'}>$100 GeV has been applied.
Initial state radiation has been considered in the structure
function approach and beamstrahlung has been generated with the aid of
the CIRCE program~\cite{circe}.
Fragmentation, hadronization and particles' decays are handled
by JETSET~\cite{jetset}, with parameters tuned to LEP2 data.

\subsection{Detector Simulation}
Once the events have been generated, they are processed through
a fast simulation~\cite{simdet} of the response of a
detector for a future linear collider. The detector components,
which are assumed to be:
\begin{itemize}
\item a vertex detector,
\item a tracker system with main tracker (2 m. radius TPC embedded in a 2 Tesla
magnetic field), 
forward tracker and forward muon tracker,
\item an electromagnetic calorimeter,
\item a hadronic calorimeter and
\item a luminosity detector,
\end{itemize}
\noindent are implemented according to the Large Detector model in~\cite{nlccdr}.

This fast detector simulation provides a flexible tool since its performance
characteristics can be varied within a wide range. The calorimeter response
is treated in a realistic way using a parametrization of the electromagnetic and
hadronic shower deposits obtained from a full GEANT simulation~\cite{bramhs} and
including a cluster finding algorithm. Pattern recognition is emulated
by means of a complete cross-reference table between generated particles and
detector response. The output of the program consists of a list of reconstructed
objects: electrons, gammas, muons, charged and neutral hadrons and unresolved
calorimeter clusters, as a result of an idealized Energy Flow (EF) algorithm 
incorporating track-cluster matching.

\subsection{B-tagging}
 Jets coming from $b$ and $c$-quark decays are tagged based on the non-zero lifetime of these
quarks, using the Vertex Detector (VDET). In this study we have assumed the performance of a 
CCD VDET in a 1 cm radius beampipe.

 In order to look for this lifetime signal, we have chosen to use the 3D impact parameter (IP) 
of each charged track (distance of closest approach between the 
track and the $b$ production point). Since the statistical resolution of the IP
varies strongly from one track to another, we use the estimated statistical significance of the 
measured IP to define our tag. The $b$-tagging algorithm is kept simple so that 
the success of the analysis
does not depend on detector details. More efficient algorithms can be developed by making use
of multivariate techniques, such as Neural Networks.

\begin{figure}[p]
\begin{center}
\mbox{
\epsfig{file=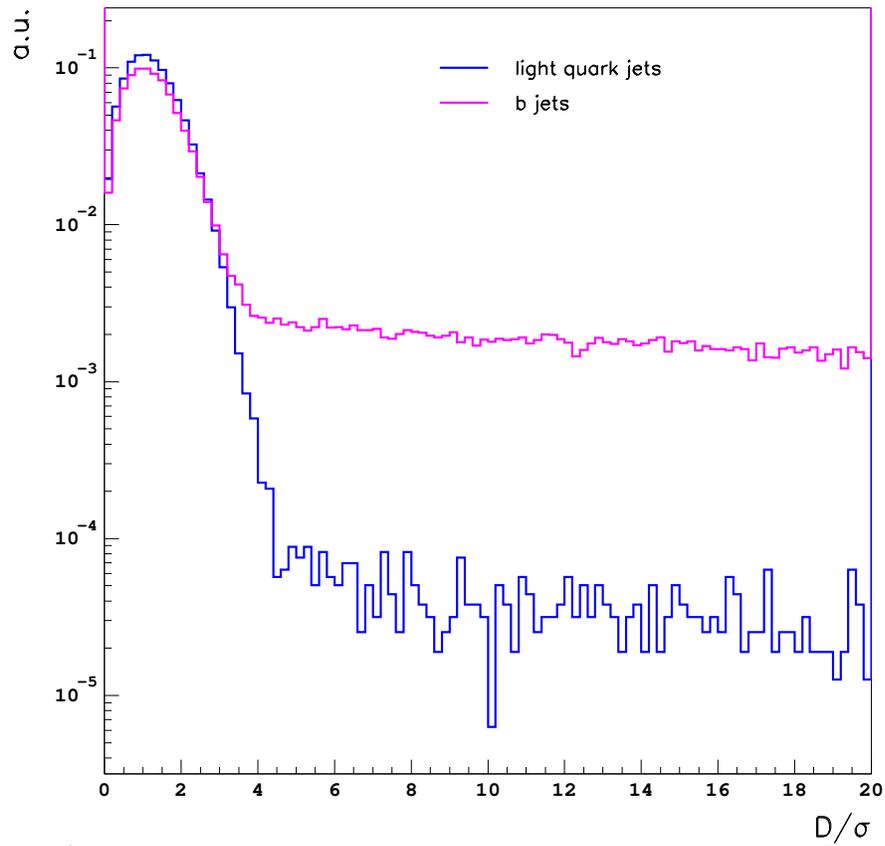,width=13cm}
}
\end{center}
\caption{\protect\footnotesize
Track 3D impact parameter significance for $Z$ hadronic events at $\sqrt{s}$=200 GeV.}
\label{btag_3dip}
\end{figure}

 In Fig.~\ref{btag_3dip}, the IP significance distributions for different $Z$ hadronic
decays are compared. The lifetime signature can be clearly seen for $Z \to b \bar{b}$ in the 
positive tail. We will use the IP distribution for prompt tracks (those originated from 
$Z \to u\bar{u},d\bar{d},s\bar{s}$)
to define, for each track, a 
probability ``to be consistent with originating from the primary vertex''. This information
can then be combined to get a probability per jet or for the whole event~\cite{aleph}. 

 In order to test the performance of such $b$-tagging, we have 
estimated its efficiency and purity for a given definition of $b$-jet. To do so, the Monte Carlo
generated quarks are assigned to the reconstructed jets by a matching algorithm which
associates those quark-jet pairs with minimum invariant mass, starting from the most energetic quark.
In order not to reduce drastically the signal efficiency, we will not use the number 
of found $b$-jets for a certain lifetime probability as a selection cut. 
Instead, for every event, we will define as 
$b$-jets those two with the lowest probability (to originate from the primary vertex). 
Applied to $ZH\to q\bar{q}b\bar{b}$ (with $q=u,d,s,c,b$) events, this algorithm would tag 
the two correct $H$ $b$-jets in $\sim43\%$ of the cases, and at least one of them in $\sim93\%$
of the cases.    

\section{Experimental Analysis}
The experimental analysis is performed assuming a 
total integrated luminosity of 10 fb$^{-1}$, which
can be collected in around 11 days of running at ${\cal L} = 10^{34}$
cm$^{-2}$s$^{-1}$.

In spite of the apparently clean signature of this decay channel (4 jets
in the final state, out of which $\geq 2$ are $b$-jets, di-jet invariant 
mass constraint for the Z decay, etc), the measurement has many difficulties, among which are:
\begin{itemize}
\item the tiny signal ($\sigma_{ZH\to q\bar{q}H} \simeq 46.2$ fb) with backgrounds about 300 times
larger: in Table~\ref{sigmas}, the total cross-sections for the
signal and different backgrounds considered are listed together with the numbers of
generated events;
\item limitations of jet-clustering algorithms in properly reconstructing
4 jets in the final state due to hard gluon radiation, jet-mixing, etc;
\item degradation of $b$-tagging performance due to hard gluon radiation;
\item missing energy in $b$ jets from neutrino emission. For instance, the
branching ratio for semileptonic+leptonic decay modes of the $B^\pm$ is $\simeq 52.4\%$. 
\end{itemize}

%%%%%%%%%%%%%%%%%%%%%%%%%%%%%%%%%%%%%%%%%%%%%%%%%%%%%%%%%%%%%%%%%%%%%%%%%%%%%%
\begin{table}[htpb]
\begin{center}
\begin{tabular}{lrc}
\hline\hline 
{\rm Process} & {$\sigma$ \rm (fb)} & {Generated events} \\
\hline
$ZH \to q\bar{q}H$ & 46.2 & 100k \\
\hline
$ZH \to \ell^+\ell^-H$ & 6.7 & 100k \\
$q\bar{q} \rm \:(5\:flavors)$ & 3860.4 & 1M\\
$t\bar{t}$ & 582.0 & 1M\\
$W^+W^-$ & 7821.4 & 2.2M\\
$ZZ$ & 570.0 & 1M\\
\hline\hline
\end{tabular}
\caption{\label{sigmas}\protect\footnotesize Total cross-section for signal and the different backgrounds
considered at $\sqrt{s}$ = 500 GeV. Initial state radiation and 
beamstrahlung have been included. For efficiency reasons, a generation cut of
$\sqrt{s'}>100$ GeV has been applied. Also listed is the number of generated
events for every process.}
\end{center}
\end{table}   
%%%%%%%%%%%%%%%%%%%%%%%%%%%%%%%%%%%%%%%%%%%%%%%%%%%%%%%%%%%%%%%%%%%%%%%%%%%%%   

Due to the very small signal-to-background (S/B) ratio, 
the philosophy of the analysis
will be to start by applying a standard-cuts preselection in order to remove as much background 
as possible while keeping a high efficiency for the signal. Then, in order to further improve
the statistical sensitivity to the signal, 
a multivariate analysis will be performed. At this stage our problem 
will be how to make an optimal use of the statistical information from a set of $N$ distributions
discriminating between signal and background. It can proven~\cite{optproj} 
that it is possible to make
an optimal projection from the input $N$-D space to a 1-D space\footnote{In the general case
of $m$ existing classes to be discriminated, the optimal projection is performed in 
a space ($m-1$)-dimensional. In our problem, all backgrounds are considered inclusively and
$m=2$, thus the optimal projection is 1-dimensional.}:
\begin{description}
\item{a)} without loss of sensitivity on the classes proportions and
\item{b)} with a probabilistic interpretation (in terms of the a-posteriori Bayesian probability
of being of signal type).
\end{description}
This projection can be performed by using Neural Network (NN) techniques, which have become
increasingly popular in High Energy Physics in the last few years.

\subsection{Event Selection}

\begin{figure}[p]
\begin{center}
\mbox{
\epsfig{file=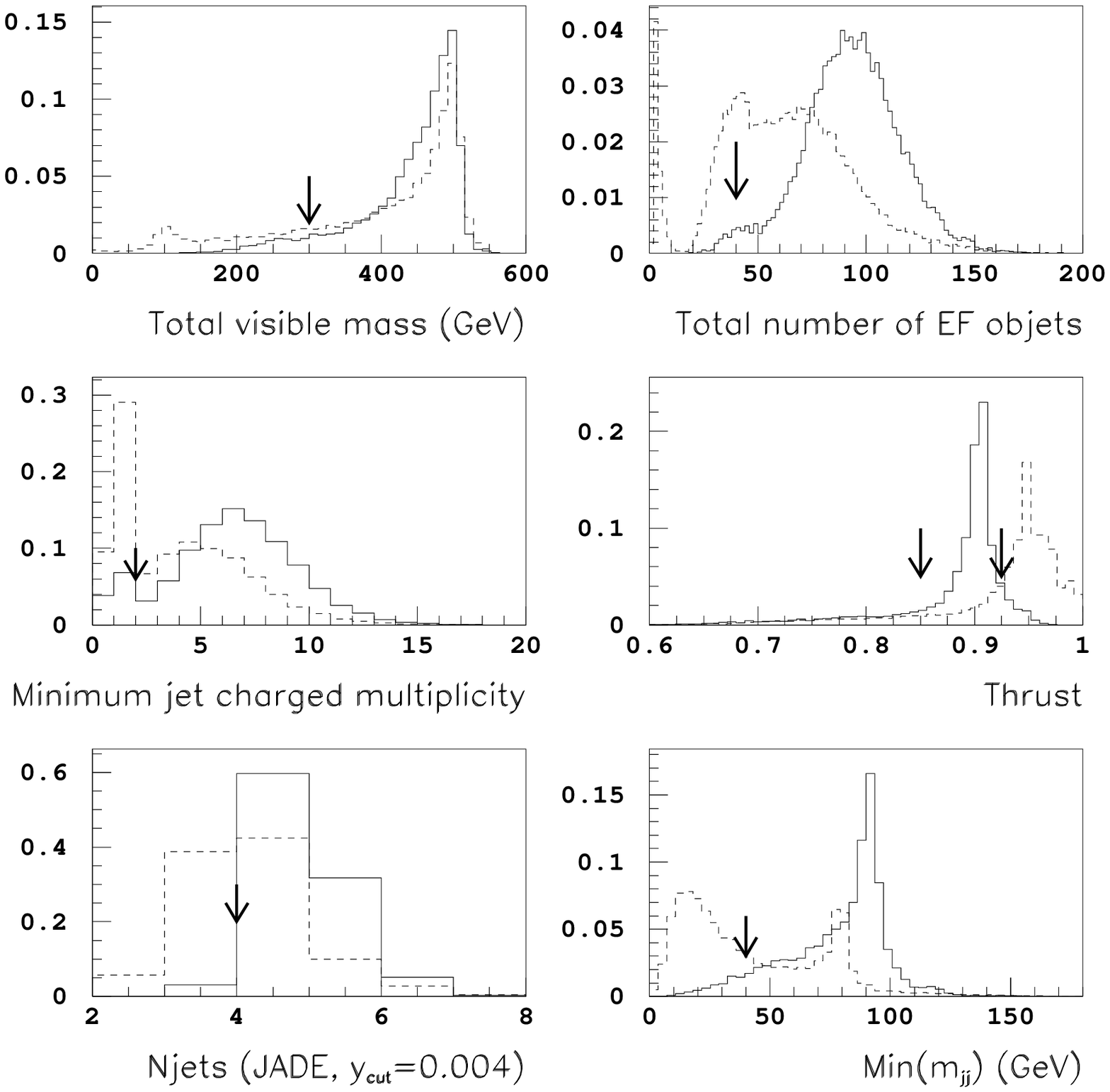,width=13cm}
}
\end{center}
\caption{\protect\footnotesize
Preselection variables for the hadronic decay channel.
Signal (solid) and background (dashed) have been normalized to the same
number of events.
The background prediction has been computed by adding all the different 
background contributions weighted according to their relative cross-section.}
\label{had_presel_cuts}
\end{figure}

As already mentioned, a standard cuts preselection is applied in order
to remove as much background as possible before the multivariate analysis.
The selected events are required to have a visible mass in excess of 0.6$\sqrt{s}$
(i.e. 300 GeV), more than 40 EF objects reconstructed,
at least 4 jets reconstructed with the JADE~\cite{JADE} jet-clustering 
algorithm with a resolution parameter
$y_{cut} = 4\times 10^{-3}$ and a thrust value ranging in between 0.85 and 0.925.
Next, the event is forced to have exactly 4 jets reconstructed using the JADE algorithm.
Further preselection cuts require a minimum of 2 charged tracks per jet and a minimum
di-jet invariant mass of 40 GeV. The preselection variables are compared
for signal and background in Fig.~\ref{had_presel_cuts}, along with the
cuts applied. The preselection efficiencies and effective
cross-sections for the different processes considered are listed in 
Table~\ref{had_pres_effs}. After preselection, the efficiency for signal 
is reduced to 67.3\% and the sample purity is only $\simeq 4.0\%$.

%%%%%%%%%%%%%%%%%%%%%%%%%%%%%%%%%%%%%%%%%%%%%%%%%%%%%%%%%%%%%%%%%%%%%%%%%%%%%%
\begin{table}[htpb]
\begin{center}
\begin{tabular}{lrr}
\hline\hline 
{\rm Process} & {$\epsilon$ (\%)} & $\sigma_{\rm eff}$ {\rm (fb)} \\
\hline
$ZH\to q\bar{q}H$ & 67.27 & 31.08 \\
\hline
$ZH\to \ell^+\ell^-H$ & 1.48 & 0.10 \\
$q\bar{q} \rm \:(5\:flavors)$ & 6.76 & 290.96\\
$t\bar{t}$ & 4.26 & 24.79\\
$W^+W^-$ & 5.00 & 391.07\\
$ZZ$ & 12.30 & 70.11\\
\hline
{\em Total Bckg} & & 747.03 \\
\hline\hline
\end{tabular}
\caption{\label{had_pres_effs}\protect\footnotesize Hadronic channel preselection efficiencies and effective cross-sections.}
\end{center}
\end{table}   
%%%%%%%%%%%%%%%%%%%%%%%%%%%%%%%%%%%%%%%%%%%%%%%%%%%%%%%%%%%%%%%%%%%%%%%%%%%%%   

\begin{figure}[p]

\begin{center}
\mbox{
\epsfig{file=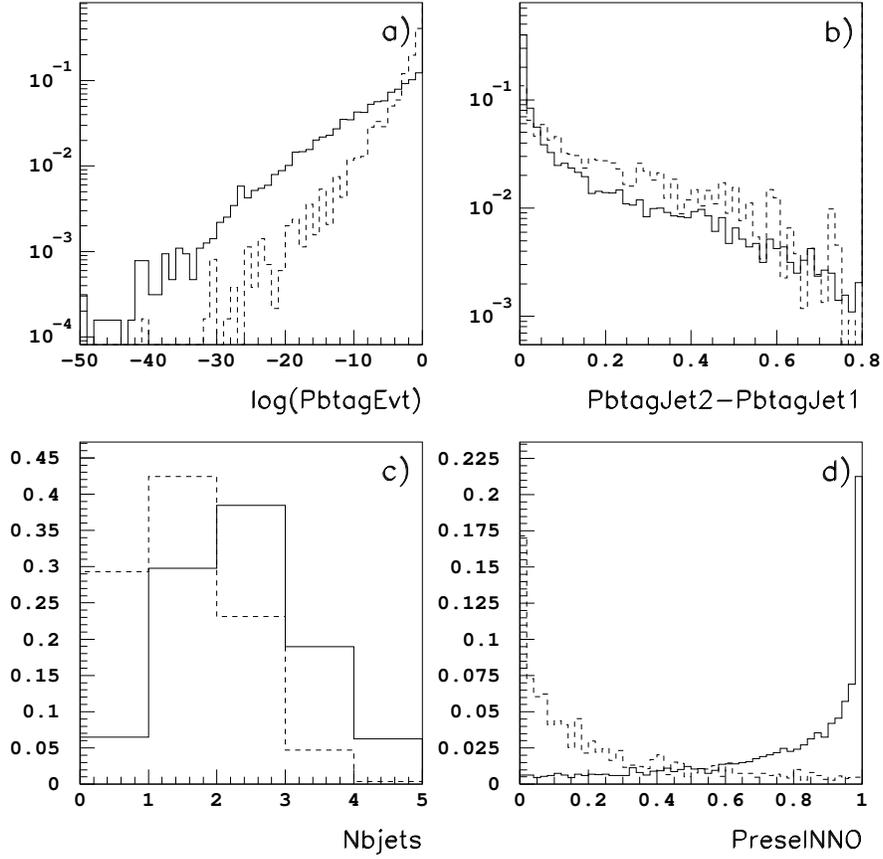,width=13cm}
}
\end{center}
\caption{\protect\footnotesize
Flavor tagging variables: a), b) and c), used together with the preselection
variables to train the Preselection NN, whose output is shown in
d). Signal (solid) and background (dashed) have been normalized to the same
number of events.
The background prediction has been computed by adding all the different 
background contributions weighted according to their relative cross-section.}
\label{had_nnpresel}
\end{figure}

As it can observed in Fig.~\ref{had_presel_cuts}, the preselection variables
after cuts still have discriminant power between signal and background. In order to
optimally use these variables, they are further used together with three more
variables to train a Preselection NN. These three variables (shown in 
Figs.~\ref{had_nnpresel}a, \ref{had_nnpresel}b and \ref{had_nnpresel}c) provide information
about the lifetime content of the event: the logarithm of the event probability
to contain no-lifetime, the difference between the probability of the second
jet and the first jet (sorted from the most $b$-like to the least $b$-like) and
the number of $b$-jets found (where a $b$-jet is defined as a jet with a no-lifetime probability
smaller than 13.5\%). In Fig.~\ref{had_nnpresel}d
it is shown the Preselection NN output, after training,
for both signal and background. No cut is applied in this distribution, 
but it is rather used as a discriminant variable.

\begin{figure}[p]
\begin{center}
\mbox{
\epsfig{file=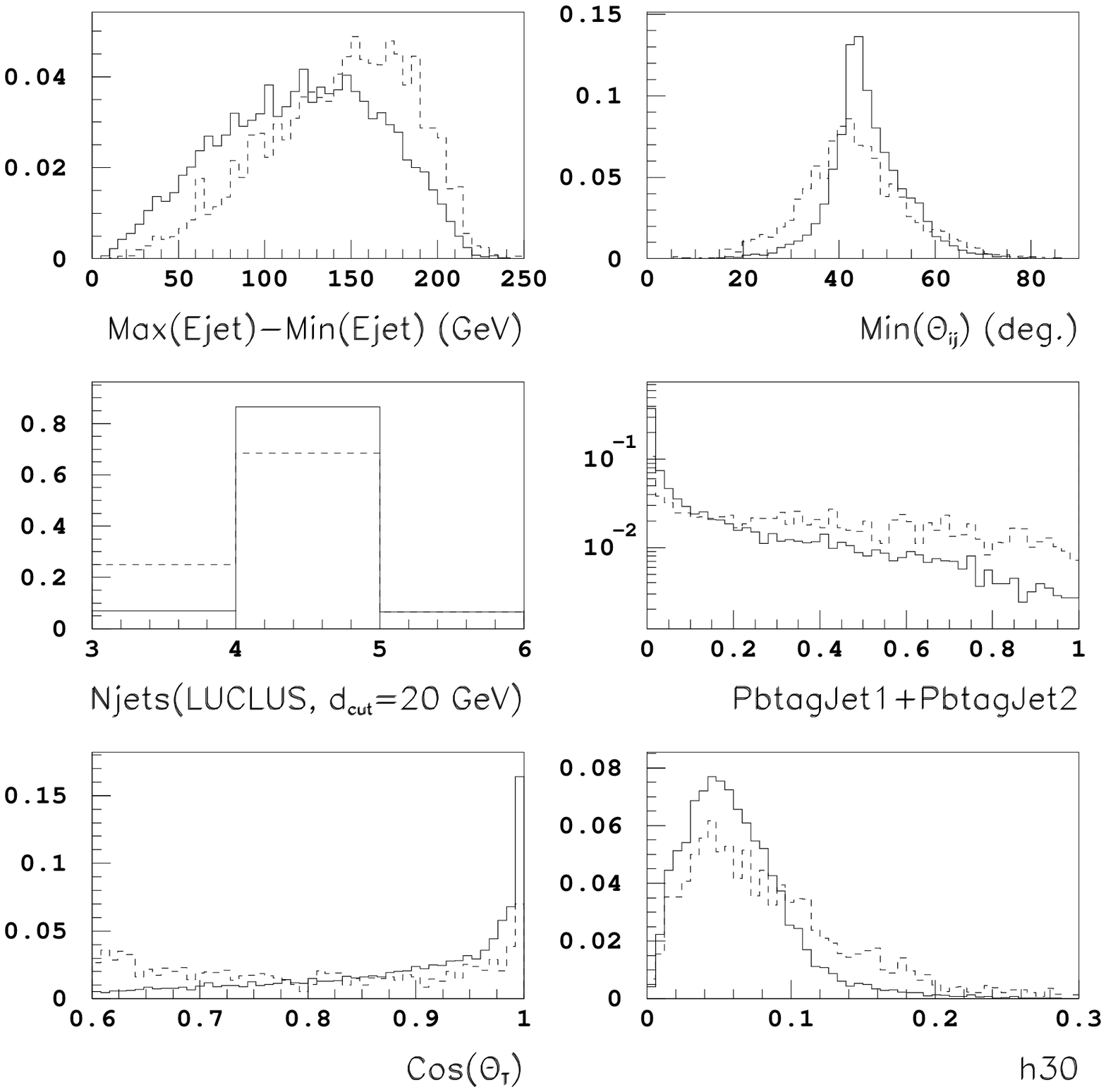,width=13cm}
}
\end{center}
\caption{\protect\footnotesize
Selection NN variables for the hadronic decay channel (I).
Signal (solid) and background (dashed) have been normalized to the same
number of events.
The background prediction has been computed by adding all the different 
background contributions weighted according to their relative cross-section.}
\label{had_nn_vars1}
\end{figure}

\begin{figure}[p]
\begin{center}
\mbox{
\epsfig{file=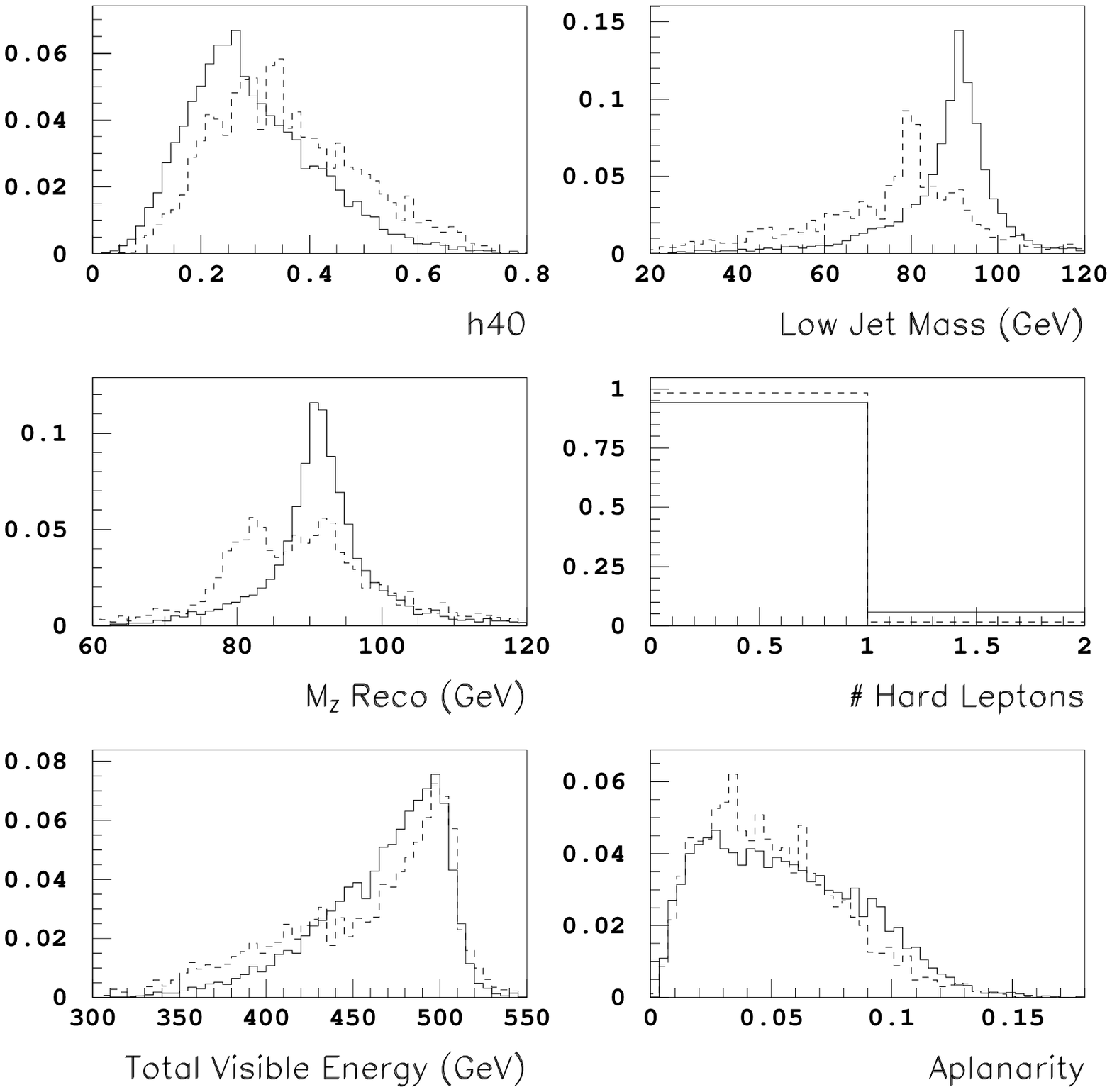,width=13cm}
}
\end{center}
\caption{\protect\footnotesize
Selection NN variables for the hadronic decay channel (II).
Signal (solid) and background (dashed) have been normalized to the same
number of events.
The background prediction has been computed by adding all the different 
background contributions weighted according to their relative cross-section.}
\label{had_nn_vars2}
\end{figure}

There are 12 more variables which are discriminant between signal and background 
(see Figs.~\ref{had_nn_vars1} and \ref{had_nn_vars2}). 
Most of them are variables about the global event topology:
\begin{itemize}
\item Evis: total visible energy of the event;\
\item Max(Ejet)-Min(Ejet): difference between maximum and minimum jet energy;
\item Njets(LUCLUS, $d_{cut}$=20 GeV): number of jets found with the LUCLUS~\cite{LUCLUS} jet-clustering
      algorithm for a distance measure of 20 GeV;
\item low jet mass of the event. The event is divided in two hemispheres and
      particles are assigned to either hemisphere in order to minimize the quadratic sum of
      the two hemispheres invariant mass (hereafter called respectively high and low jet masses). 
      For processes with two resonances (such as $W^+W^-$, $ZZ$ or $ZH$),
      these distributions tend to show resonant structures around the true invariant masses.
\item minimum di-jet angle;
\item cosine of the polar angle of the thrust axis;
\item normalized Fox-Wolfram moments $h_{30}$ and $h_{40}$;
\item aplanarity,
\item number of hard leptons ($E>50$ GeV) found;
\end{itemize}
\noindent others contain information about flavor tagging (sum of the no-lifetime probability 
for the two most $b$-like jets) or $Z$ invariant mass reconstruction (see Sect.~\ref{hmreco}).

\begin{figure}[p]
\begin{center}
\mbox{
\epsfig{file=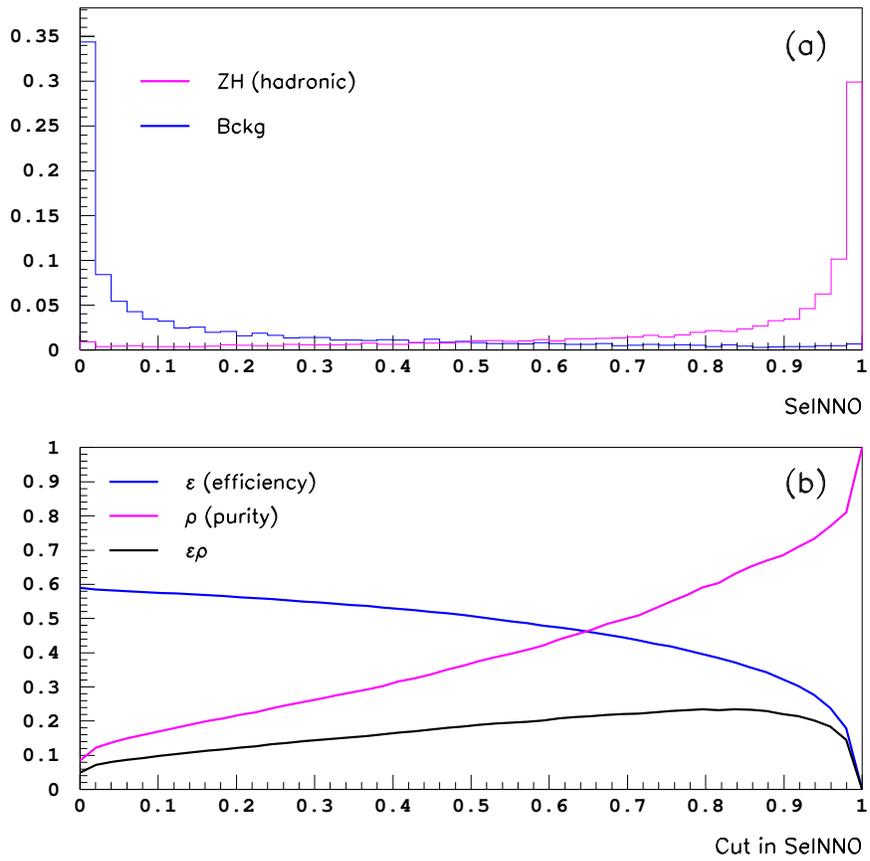,width=13cm}
}
\end{center}
\caption{\protect\footnotesize
(a) Selection NN output.
Signal (purple) and background (blue) have been normalized to the same
number of events. (b) Efficiency and purity as a function of the cut
in the Selection NN output.}
\label{had_nnout_sel}
\end{figure}

\begin{figure}[p]
\begin{center}
\mbox{
\epsfig{file=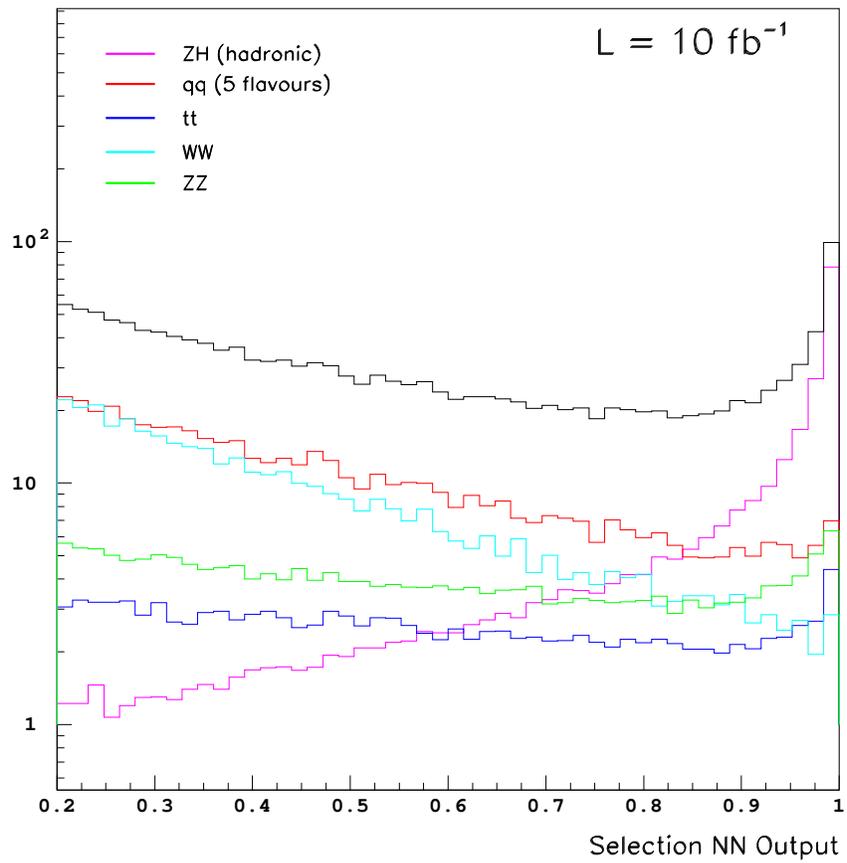,width=13cm}
}
\end{center}
\caption{\protect\footnotesize
Selection NN output. For the sake of clarity, the NN output is restricted
to be larger than 0.2. The different contributions have been
normalized to the same integrated luminosity.}
\label{had_nnout_sel_color}
\end{figure}

These variables, together with the Preselection NN output (PreselNNO) distribution are used
to train a Selection NN. Table~\ref{had_selnn_discpower} shows the discriminant power
of each of the 13 variables used in the Selection NN.
The Selection NN output (SelNNO) distribution is
compared for signal and background in Fig.~\ref{had_nnout_sel}a. In Fig.~\ref{had_nnout_sel}b,
the signal efficiency ($\epsilon$) and purity ($\rho$) as a function of the cut in the SelNNO are shown.
Among the different backgrounds, the main contribution in the ``signal region''
(e.g. SelNNO$>0.85$) comes from $q\bar{q}$ (5 flavors), followed by
$ZZ$, as shown in Fig.~\ref{had_nnout_sel_color}. 

%%%%%%%%%%%%%%%%%%%%%%%%%%%%%%%%%%%%%%%%%%%%%%%%%%%%%%%%%%%%%%%%%%%%%%%%%%%%%%
\begin{table}[htpb]
\begin{center}
\begin{tabular}{lc}
\hline\hline 
{\rm Variable} & {\rm Discriminant Power (\%)} \\
\hline
${\rm max(E^{jet})-min(E^{jet})}$             &  8.8  \\
${\rm min(\theta_{ij})}$                      &  7.6  \\
${\rm N_{jets} (LUCLUS)}$                     &  5.2 \\ 
${\rm P_{btagOrd}^{jet1}+P_{btagOrd}^{jet2}}$ &  4.1\\
${\rm cos(\Theta_T)}$                         &  6.0\\
${\rm h_{30}}$                                & 11.0\\
${\rm h_{40}}$                                &  9.1\\
Low jet mass                                  &  8.2  \\
${\rm M_Z^{reco}}$                            &  9.8 \\
Number of hard leptons                        &  4.3  \\
${\rm E_{vis}}$                               &  9.9 \\
Aplanarity                                    &  7.2 \\
PreselNNO                                     &  8.7 \\ 
\hline\hline
\end{tabular}
\caption{\label{had_selnn_discpower}\protect\footnotesize 
Discriminant power of each of the 12 input variables of the Selection NN.} 
\end{center}
\end{table}   
%%%%%%%%%%%%%%%%%%%%%%%%%%%%%%%%%%%%%%%%%%%%%%%%%%%%%%%%%%%%%%%%%%%%%%%%%%%%%
The selection can be performed in such a way as to minimize the statistical uncertainty in
the cross-section measurement and thus, in the Z-Higgs Yukawa coupling:
\begin{eqnarray}
\biggl (\frac{\Delta g_{ZZH}}{g_{ZZH}} \biggr)_{stat} = 
\frac{1}{2} \biggl (\frac{\Delta \sigma_{ZH}}{\sigma_{ZH}} \biggr)_{stat} =
\frac{1}{2} \frac{1}{\sqrt{\sigma_{ZH\to q\bar{q}H}}}\frac{1}{\sqrt{\epsilon \rho L}}, \nonumber 
\end{eqnarray}
\noindent where $L$ is the integrated luminosity. 
This cut would correspond to the maximum of $\epsilon\rho$ shown
in Fig.~\ref{had_nnout_sel}b, $(\epsilon\rho)_{max} \simeq 0.23$,
which translates into:
\begin{eqnarray}
\biggl (\frac{\Delta \sigma_{ZH}}{\sigma_{ZH}} \biggr)_{stat} \simeq 9.7\%,\quad
\biggl (\frac{\Delta g_{ZZH}}{g_{ZZH}} \biggr)_{stat} \simeq 4.9\%, \nonumber
\end{eqnarray}
\noindent assuming 10 fb$^{-1}$ of integrated luminosity\footnote{A better statistical
uncertainty could in principle be obtained from a likelihood fit to the SelNNO distribution.}.
This optimal cut, SelNNO$>$0.85, 
leads to a signal efficiency of 38.0\% and a purity of 56.2\%.
The selection efficiencies and effective cross-sections for the different
backgrounds are given in Table~\ref{had_Pres_Effs}, corresponding to the above cut.
However, in order to determine the Higgs mass, a higher sample purity is in general
desirable, which can be obtained by performing a tighter cut on SelNNO.

%%%%%%%%%%%%%%%%%%%%%%%%%%%%%%%%%%%%%%%%%%%%%%%%%%%%%%%%%%%%%%%%%%%%%%%%%%%%%%
\begin{table}[htpb]
\begin{center}
\begin{tabular}{lrcc}
\hline\hline 
{\rm Process} & {$\epsilon$ (\%)} & $\sigma_{\rm eff}$ {\rm (fb)} & {\rm \# Events (L=10 fb$^{-1}$)}\\
\hline
$ZH\to q\bar{q}H$ & 38.0  & 17.56 & 176\\
\hline
$ZH\to \ell^+\ell^-H$ & $\sim 0$ & $\sim 0$ & $\sim 0$\\
$q\bar{q} \rm \:(5\:flavors)$ & $1.30\times 10^{-1}$ & 5.02 & 50\\
$t\bar{t}$ & $3.92\times 10^{-1}$ & 2.28 & 23\\
$W^+W^-$ & $3.40\times 10^{-2}$ & 2.66 & 27\\
$ZZ$ & $6.50\times 10^{-1}$ & 3.70 & 37\\
\hline
{\em Total Bckg} & & 13.66 & 137 \\
\hline\hline
\end{tabular}
\caption{\label{had_Pres_Effs}\protect\footnotesize Selection efficiencies 
and effective cross-sections for SelNNO$>$0.85.}
\end{center}
\end{table}   
%%%%%%%%%%%%%%%%%%%%%%%%%%%%%%%%%%%%%%%%%%%%%%%%%%%%%%%%%%%%%%%%%%%%%%%%%%%%%   

\subsection{Higgs mass reconstruction}
\label{hmreco}

Once the events have been selected, the Higgs invariant mass has to be
reconstructed. In this analysis we are not going to be exclusive in the reconstruction
of the different Higgs decay channels,
but rather try to reconstruct always 4 jets in the final state. Indeed, this leads to some
inefficiency (in particular for $H$ decay modes such as $H\to W^+W^-,\tau^+\tau^-$),
but since $BR(H\to b\bar{b}+c\bar{c}+gg) \simeq 80\%$, it is fully justified for
the purpose of this study. 

For four reconstructed jets in the final state there are 6
possible di-jet assignments. At this point we do not use the $b$-tagging information
in order to identify the $b$-jets coming from the Higgs, nor any assumption about the
Higgs mass. Instead, the combination which 
maximizes ${\cal P}(m_{ij}-m_Z)$, where $m_{ij}$ is the invariant mass between
jets $i$ and $j$, and ${\cal P}$ is the 
probability density function of the $Z$ invariant mass for the correct jet pairing, is selected. 
The efficiency to tag the correct combination is 68\% before preselection 
and goes up to 85\% for the finally selected events. Therefore, the combinatorial background
in the final Higgs invariant mass distribution is $\sim 15\%$.

\begin{figure}[htbp]
\begin{center}
\mbox{
\epsfig{file=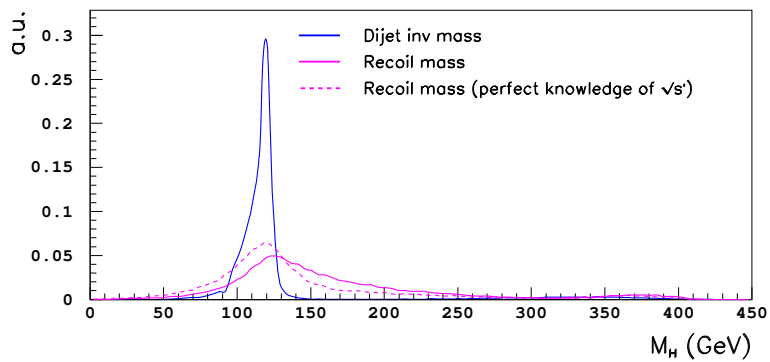,width=11cm}
}
\end{center}
\caption{\protect\footnotesize
Comparison between the raw $H$ di-jet invariant mass and
recoil mass distributions. A selection cut of SelNNO$>$0.9 has been
applied but no background has been included. 
All distributions have been normalized to the same number of events.}
\label{recoil}
\end{figure}

\begin{figure}[htbp]
\begin{center}
\mbox{
\epsfig{file=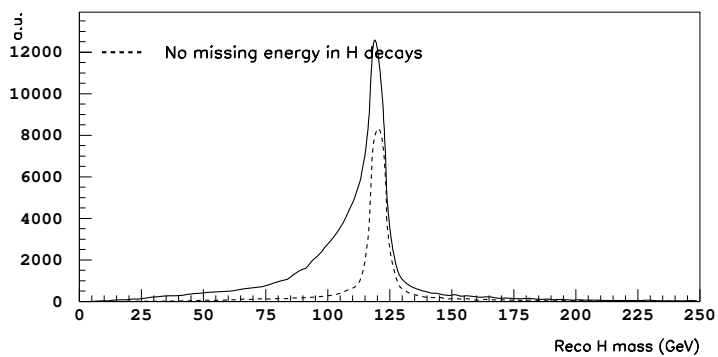,width=11cm}
}
\end{center}
\caption{\protect\footnotesize
Raw $H$ di-jet invariant mass distribution. The contribution from
events with missing energy below 2\% shows a sharp and symmetric
distribution around 120 GeV (dashed). No background has been included.}
\label{mhdistrib_inclusive}
\end{figure}

As already mentioned, the raw recoil mass distribution from the reconstructed $Z$
jets is not really suited for a precise $M_H$ measurement in the hadronic channel, 
even assuming perfect knowledge
of the event-by-event effective center-of-mass energy. Instead, the raw di-jet invariant mass
distribution from the $H$ jets shows a much sharper peak around 120 GeV, as shown
in Fig.~\ref{recoil}. The asymmetry and low mass tail for the the raw $H$ di-jet invariant mass
distribution is caused by energy losses in the $H$ decays (dominated by neutrino
emission in $b$ decays from $H\to b\bar{b}$, but also receiving a small contribution from
other $H$ decay modes such as $\tau^+\tau^-$ or $W^+W^-$). This is demonstrated  
in Fig.~\ref{mhdistrib_inclusive}, where the amount of energy 
lost in the form of neutrinos by each $H$ jet has been computed at the hadron level from the
MC. The fraction of energy lost is defined for each $H$ jet with respect to the 
true $H$ daughter's energy. Then, the sum of both fractions is required
to be below 2\% for an event to be considered with no missing energy in the $H$ decay.

\begin{figure}[p]
\begin{center}
\mbox{
\epsfig{file=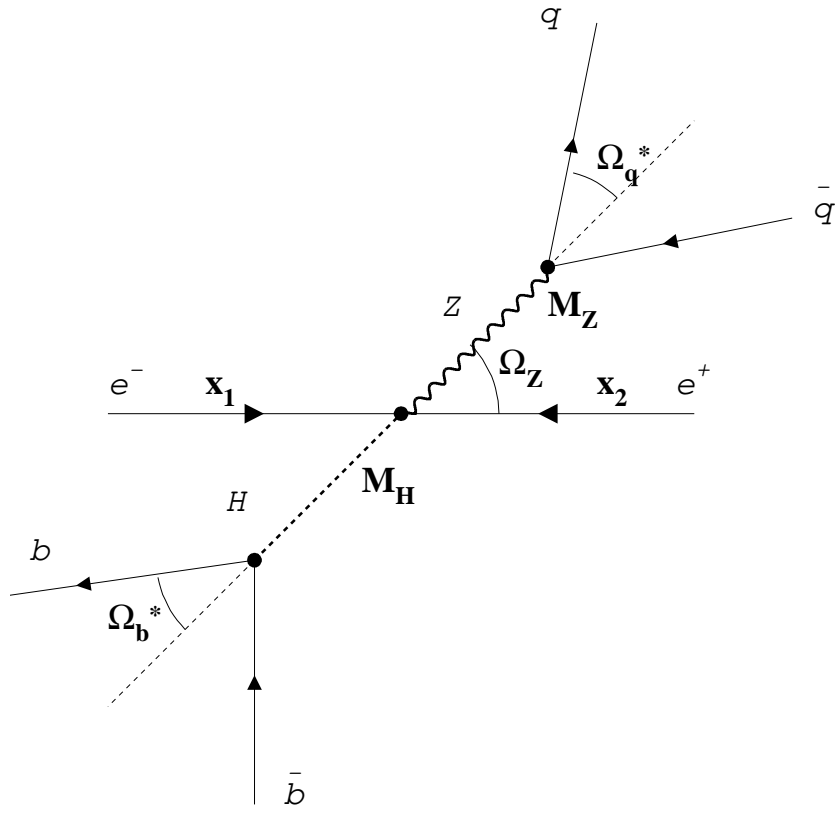,width=13cm}
}
\end{center}
\caption{\protect\footnotesize
Set of kinematical variables to describe $ZH$ production. The angular variables
are generically denoted by $\Omega = (\theta,\phi)$.}
\label{diagram}
\end{figure}

In order to improve the Higgs invariant mass resolution, a kinematical fit (KF)
imposing energy and momentum conservation is performed. In an event-by-event basis,
the whole event kinematics (represented in Fig.~\ref{diagram}) is fitted:
\begin{itemize}
\item di-jet invariant masses: $M_Z$ and $M_H$;
\item production angles of the $Z$ boson in the
$e^+e^-$ rest frame: $\theta_Z$ and $\phi_Z$;
\item production angles of one of the $Z$ jets with respect to the $Z$ 
direction in the $Z$ rest frame: $\theta_q^*$ and $\phi_q^*$;
\item production angles of one of the $H$ jets with respect to the $H$ 
direction in the $H$ rest frame: $\theta_b^*$ and $\phi_b^*$.
\end{itemize}

\begin{figure}[p]
\begin{center}
\mbox{
\epsfig{file=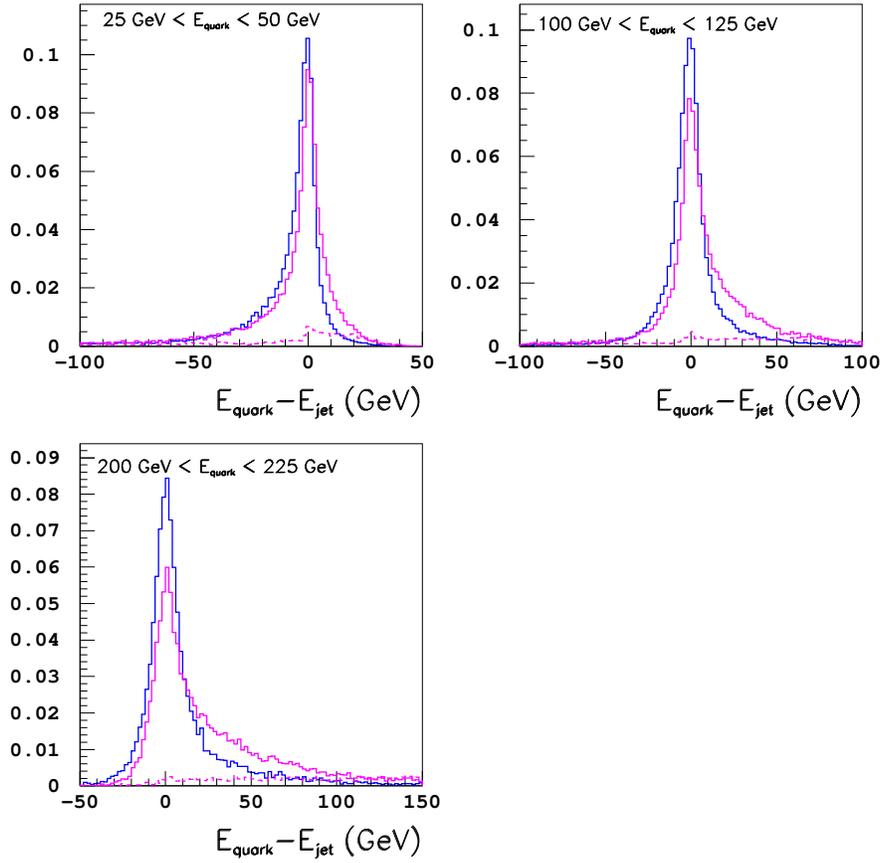,width=13cm}
}
\end{center}
\caption{\protect\footnotesize
Jet energy resolution for light-quark jets (from $Z$ decays, blue solid histogram) and
``$b$-jets'' (from $H$ decays, magenta solid histogram) in different parent quark energy 
ranges. The contribution from $H\to W^+W^-,\tau^+\tau^-$ decay channels
is also explicited (magenta dashed histogram).}
\label{jetenedistrib}
\end{figure}

The jet masses have been fixed to the reconstructed values.
It order to properly correct the jet energies and angles, we have included
in the kinematical fit the non-gaussian probability density functions (PDFs):
\begin{eqnarray}
f(E_q-E_j\mid E_q),\quad f(\theta_q-\theta_j\mid E_q),\quad f(\phi_q-\phi_j\mid E_q), \nonumber
\end{eqnarray}
\noindent where $E_{q(j)}$, $\theta_{q(j)}$ and $\phi_{q(j)}$ are respectively the
energy and angles of the quarks(jets) in the laboratory frame. This is particularly
important for $b$-jets, as can be observed in Fig.~\ref{jetenedistrib}. Therefore, the
above PDFs have been parametrized separately for light-quark and $b$-jets.

\begin{figure}[p]
\begin{center}
\mbox{
\epsfig{file=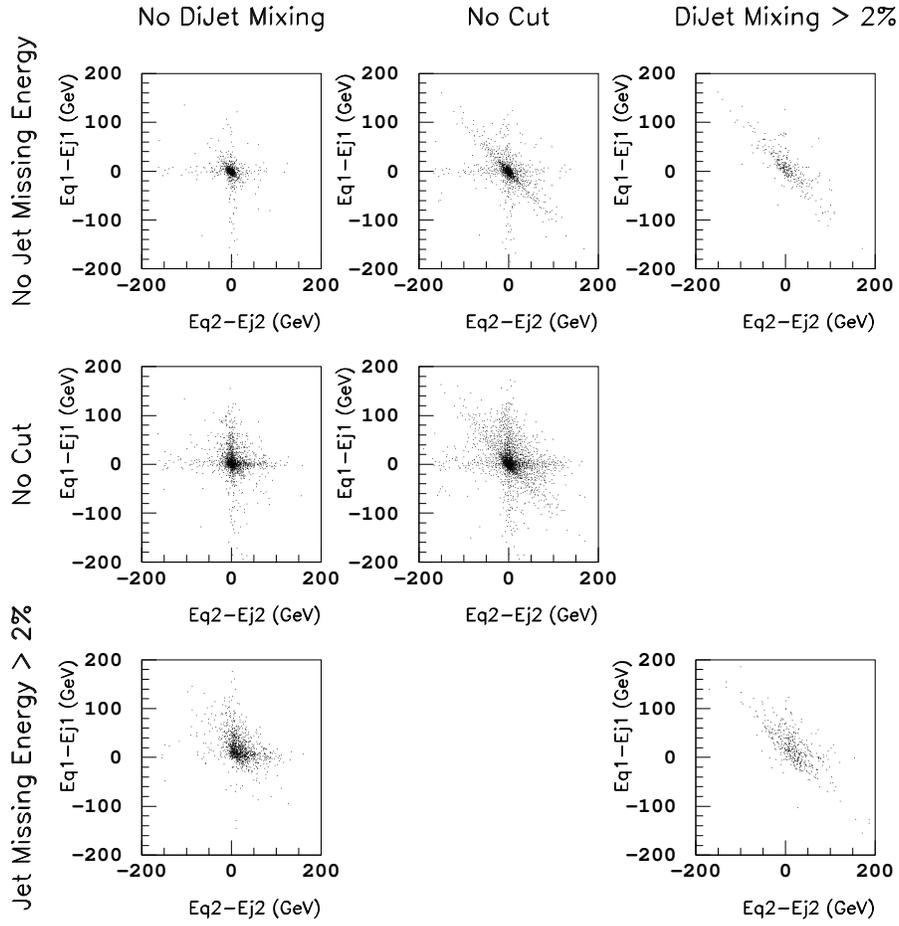,width=13cm}
}
\end{center}
\caption{\protect\footnotesize
Effect of jet-mixing and energy losses on the bi-dimensional distribution
of energy resolution for both $H$ jets. Only 
$H\to b\bar{b},c\bar{c},gg$ decay channels have been included.}
\label{eureka}
\end{figure}

In Fig.~\ref{jetenedistrib}, the jet energy resolution is compared for light-quark
jets from the $Z$  and ``$b$-jets''\footnote{As already mentioned, we have
been inclusive in the treatment of $H$ decay modes other than $b\bar{b}$ and they are included
in the histograms.} from the $H$ 
in 3 different quark energy ranges. The contribution
from $H\to W^+W^-, \tau^+\tau^-$ has also been explicited.
As it can be observed, at low parent quark energy the
jet energy resolution distribution is very similar for both $Z$ light-quark and
$H$ $b$-jets and shows a negative tail because of jet-mixing with the
other jet from the same boson decay, whereas the mixing between jets belonging to different
boson decay is negligible. The main reason is the large boost of the $Z$ and $H$, which
reduces the angular separation between the decay products belonging to the same boson:
\begin{eqnarray}
<\theta_{jet\in H(Z),jet\in H(Z)}> &\simeq& 70^{\rm o}(56^{\rm o}), \nonumber \\
<min(\theta_{jet\in Z,jet\in H})> &\simeq& 120^{\rm o}. \nonumber
\end{eqnarray}
As the parent quark energy increases, the difference between the jet energy resolution
for light-quark and $b$-jets becomes more evident, the latter developing a 
larger positive tail. The reconstructed jet energy is lower than the quark energy
because of losses in neutrino emission, which are not however larger for jets coming from
lower energy quarks, but which become more evident at high energy because of the better
performance of the jet-clustering algorithm in properly reconstructing the jet. 
Fig.~\ref{eureka} shows the effect of jet-mixing and missing energy on the
bi-dimensional distribution of energy resolution for both $H$ jets (only 
$H\to b\bar{b},c\bar{c},gg$ decay channels have been included). As expected, jet energy losses
are uncorrelated for both $H$ jets, whereas jet-mixing introduces a clear anticorrelation.
The jet-mixing is computed at the hadron level and defined as the fraction of the reconstructed
jet energy coming for the other quark in the same boson decay.

\begin{figure}[p]
\begin{center}
\mbox{
\epsfig{file=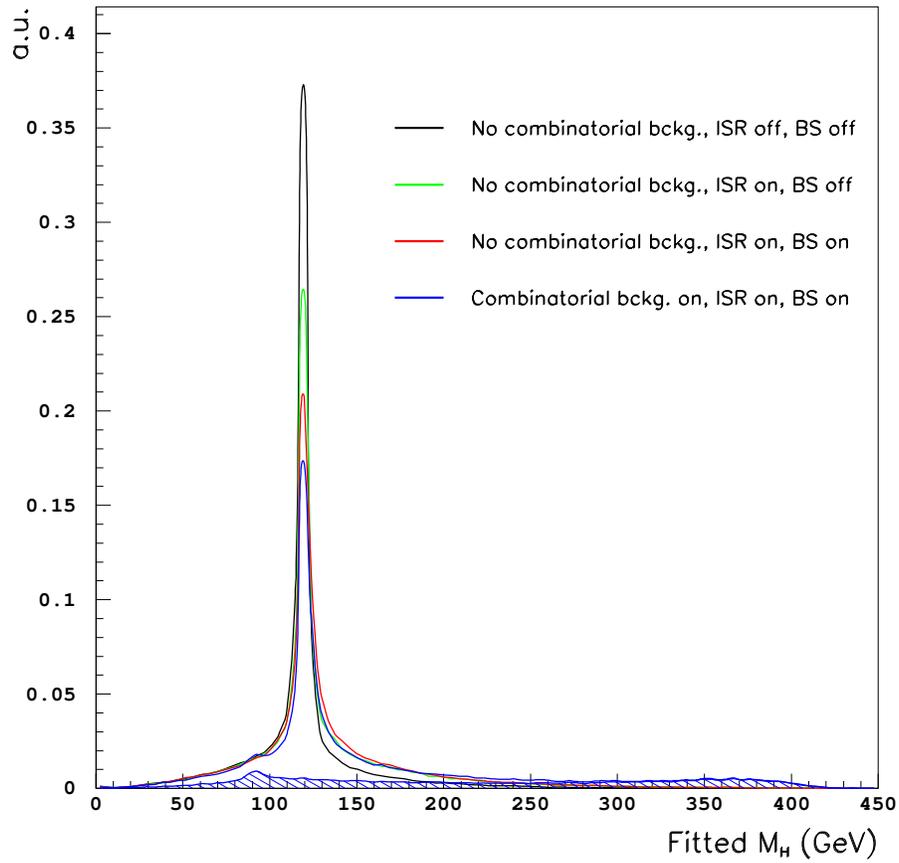,width=13cm}
}
\end{center}
\caption{\protect\footnotesize
Effect of initial state radiation (ISR), beamstrahlung (BS) and combinatorial
background on the fitted Higgs invariant mass. ISR and BS are not taken into account
in the kinematical fit. The shaded histogram shows the contribution from
the combinatorial background. All distributions have been normalized to the same
number of events.}
\label{fittedmh_vs_isrbs}
\end{figure}

\begin{figure}[htbp]
\begin{center}
\mbox{
\epsfig{file=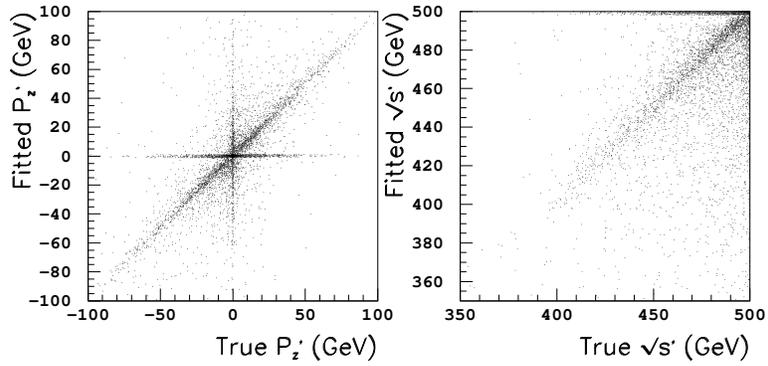,width=11cm}
}
\end{center}
\caption{\protect\footnotesize
True versus fitted total longitudinal momentum (left) and effective center-of-mass
energy (right).}
\label{fitisr}
\end{figure}

\begin{figure}[htbp]
\begin{center}
\mbox{
\epsfig{file=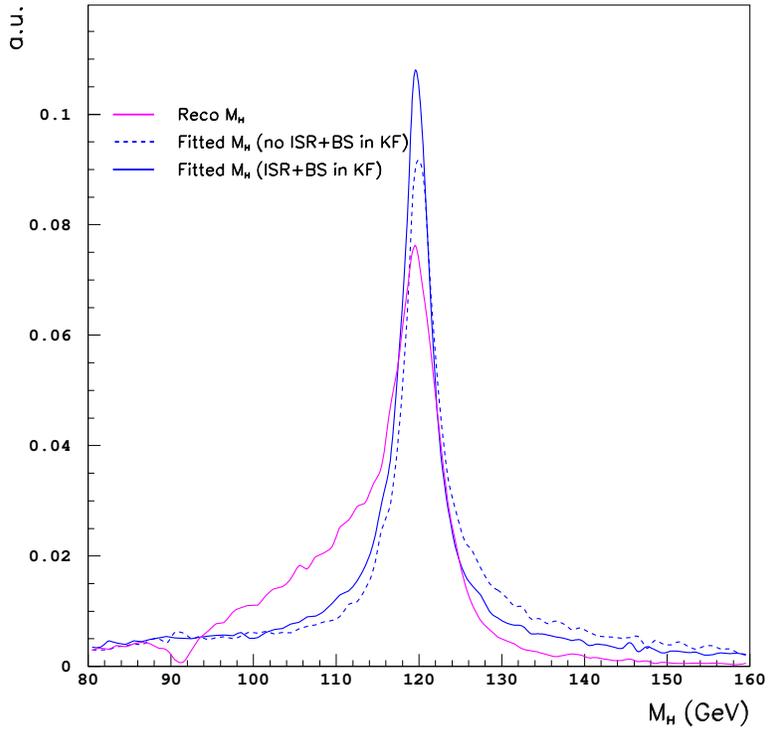,width=11cm}
}
\end{center}
\caption{\protect\footnotesize
Comparison between raw reconstructed (magenta) and fitted Higgs invariant
mass distribution (blue). The improvement by including ISR and BS
in the kinematical fit is clearly observed. No background has been included.}
\label{nofitisrvsfitisr}
\end{figure}

In order to further improve the $H$ invariant mass resolution, it is necessary to properly account
for initial state radiation and beamstrahlung in the kinematical fit. As it can
be observed in Fig.~\ref{fittedmh_vs_isrbs}, it constitutes the second largest source 
of degradation in the $H$ di-jet invariant mass resolution 
(the first one being energy losses in $b$-jet decays). In order to account for the
event-by-event fluctuations in the effective center-of-mass energy and boost
along the $z$-direction, the 
fraction of energy carried by the $e^-$ and the $e^+$: $x_1$ and $x_2$,
after ISR and beamstrahlung is fitted by including in
the likelihood the ISR structure functions for both the
$e^-$ and $e^+$. Indeed, it would be best to include the effective 
structure functions (after selection) accounting for ISR and beamstrahlung.
However, even this simple approach gives good results, as it is shown in
Fig.~\ref{fitisr}, where a clear correlation between the true and fitted total
longitudinal momentum and the true and fitted effective center-of-mass energy is observed.
Fig.~\ref{nofitisrvsfitisr} illustrates the overall improvement in the
$H$ invariant mass resolution by performing a kinematical fit with respect to
the raw reconstructed di-jet invariant mass, as well as the further gain
obtained by including ISR and beamstrahlung in the kinematical fit.

\subsection{Higgs mass determination}

\begin{figure}[p]
\begin{center}
\mbox{
\epsfig{file=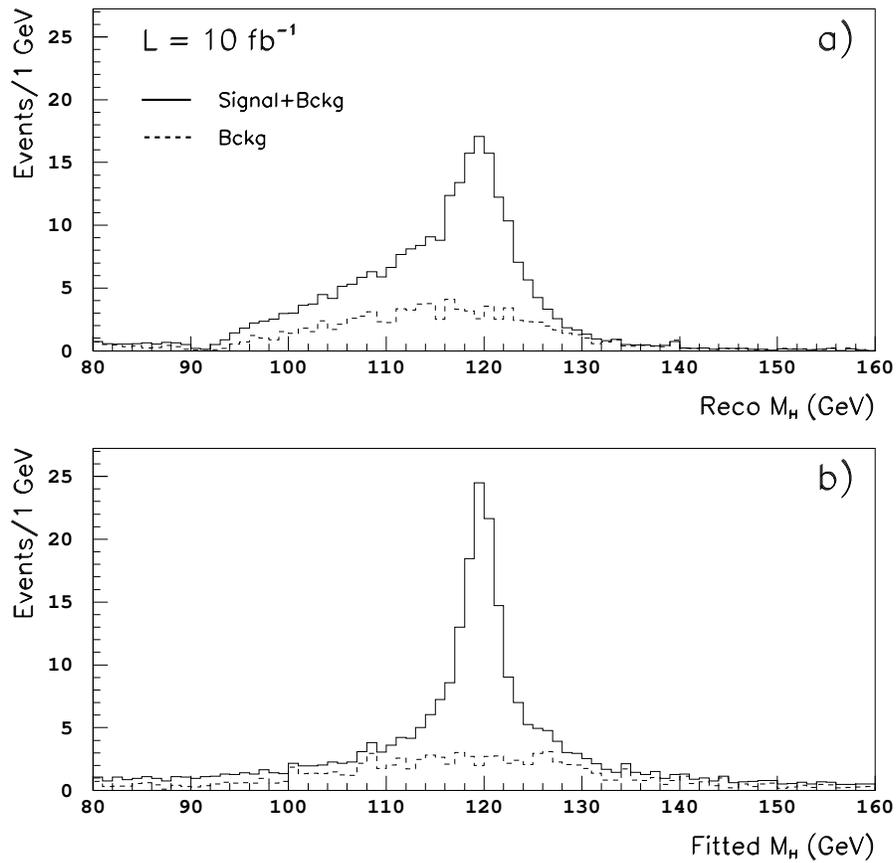,width=13cm}
}
\end{center}
\caption{\protect\footnotesize
Higgs invariant mass distribution corresponding to $L = 10$ fb$^{-1}$
and including background (dashed): (a) raw reconstructed di-jet invariant mass, (b)
di-jet invariant mass from the kinematical fit taking into account ISR and beamstrahlung.}
\label{fittedmh_final}
\end{figure}

The Higgs mass is determined from a likelihood fit to the $H$ 
invariant mass distribution resulting from the kinematical fit ($M_H^{KF}$).
The data sample corresponds to $\int {\cal L}dt=10$ fb$^{-1}$
and includes background (see Fig.~\ref{fittedmh_final}b). 
The Higgs mass estimator for a particular data sample
containing $N_{data}$ events, 
$\hat{M}_H$, is obtained by maximizing the log-likelihood function:
\begin{eqnarray}
{\cal L}(M_H) = -2\log L(M_H) = -2 \sum^{N_{data}} 
\log \left \{ \rho\:{\cal P}_S(M_H^{KF} \mid M_H) + (1-\rho)\:{\cal P}_B(M_H^{KF}) \right \}
\end{eqnarray}
\noindent where $\rho$ is the expected signal purity and
${\cal P}_{S(B)}$ is the signal (background) PDF. 
The fit is performed in the range 115 GeV $\leq M_H^{KF} \leq$ 125 GeV, where
most of the sensitivity to $M_H$ exists and the signal PDF is to a good
approximation a truncated Breit-Wigner distribution with $\sim$ 2.2 GeV width.
Apart from the cut in the $H$ invariant mass distribution, a
selection cut SelNNO$>$0.9 has been applied, leading to a signal efficiency 
of 16.5\% and a purity of 82.1\%. The expected numbers of signal and background events
in the data sample are 76.2 and 13.6, respectively.

A number of MC experiments is performed and the expected error in the
Higgs mass is computed as the RMS of the distribution of $M_H$ estimators.
The $M_H$ estimator used has been checked to be unbiased by comparing the mean of the
distribution of estimators with the input Higgs mass.
The resulting statistical uncertainty on $M_H$ is:
\begin{eqnarray}
(\Delta \hat{M}_H)_{stat} \simeq 350\: {\rm MeV}, \nonumber
\end{eqnarray}
\noindent corresponding to $\int {\cal L}dt=10$ fb$^{-1}$.
The effect of the background contamination is a $\sim 7\%$ degradation in the
statistical uncertainty. The effect of taking into account ISR
and beamstrahlung in the kinematical fit has been a $\sim 20\%$ decrease in the 
statistical uncertainty.

\section{Conclusions}
In this work we have focussed on the feasibility of a precise measurement
of the mass of a 120 GeV MSM Higgs boson by direct reconstruction, that would be
attained at a high luminosity $e^+e^-$ future linear collider operating
at a center-of-mass energy of $\sqrt{s}=500$ GeV. 

In common with previous studies, we have considered the Higgstrahlung
process: $e^+e^- \to Z^*\to ZH$, which constitutes the main Higgs
production mechanism at $\sqrt{s}\leq 500$ GeV for $M_Z \leq M_H \leq 2 M_W$.
However, most results found in the literature have focussed on the
$Z$ leptonic decay channel: $ZH \to \ell^+\ell^- H$, due to the smaller
background, the better intrinsic resolution on the Higgs invariant mass
through the recoil mass distribution and the possibility of a
measurement independent of assumptions about the Higgs decay modes.

Here we have rather concentrated on the $Z$ hadronic decay channel:
$ZH \to q\bar{q}H$, which has the bonus of a much larger statistics,
but also the complications associated with a larger background,
limitations of jet-clustering algorithms in properly reconstructing
multi-jet final states, poor di-jet invariant mass resolution, etc.
Much effort has been put in performing a ``realistic simulation''
by including irreducible and reducible backgrounds as well as
realistic detector effects and reconstruction procedures.
In order to fully exploit the possibilities of this decay channel, the use
of sophisticated tools such as Neural Networks and kinematical fitting
has been found to be important. As a result, the Higgs mass
and Z-Higgs Yukawa coupling can be determined with a statistical
accuracy exceeding that of the $Z$ leptonic decay channel.
For illustrative purposes only, the results of a recent study~\cite{pablo}
making use of $ZH \to \ell^+\ell^-H,\: \ell=e,\mu$ events at $\sqrt{s}=350$ GeV
have been naively rescaled to $\sqrt{s}=500$ GeV and compared to the
results from this work:

\begin{eqnarray}
ZH \to \ell^+\ell^-H,\: \ell=e,\mu: &&(\Delta M_H)_{stat} \simeq 160\:{\rm MeV}, 
\quad \biggl (\frac{\Delta g_{ZZH}}{g_{ZZH}}\biggr )_{stat} \simeq 3.8\%, \nonumber \\
ZH \to q\bar{q}H: &&(\Delta M_H)_{stat} \simeq 50\:{\rm MeV}, 
\quad \biggl (\frac{\Delta g_{ZZH}}{g_{ZZH}}\biggr )_{stat} \simeq 0.7\%, \nonumber 
\end{eqnarray}

\noindent assuming $\int {\cal L}dt=500$ fb$^{-1}$ of integrated luminosity. 

\section{Acknowledgements}
The author would like to thank the organizers of the LCWS99
meeting at Sitges for the pleasant setting and stimulating atmosphere
provided. The author is also grateful to Dr. Hugh E. Montgomery for 
reading this manuscript and the constructive criticism provided.
This work has been supported by the U.S. Department of Energy
under contract number DE-AC02-76CH03000.

\newpage
%%%%%%%%%%%%%%%%%%%%%%%%%%%%%%%%%%%%%%%%%%%%%%%%%%%%%%%%%%%%%%%%%%%%%%%%%%%%
%%%%%%%%%%%%%%%%%%%%%%%%%%%%%%%%%%%%%%%%%%%%%%%%%%%%%%%%%%%%%%%%%%%%%%%%%%%%
%%%%%%%%%%%%%%%%%%%%%%%%%%%%%%%%%%%%%%%%%%%%%%%%%%%%%%%%%%%%%%%%%%%%%%%%%%%%
\begin{figure}[htbp]
\begin{center}
\mbox{
\epsfig{file=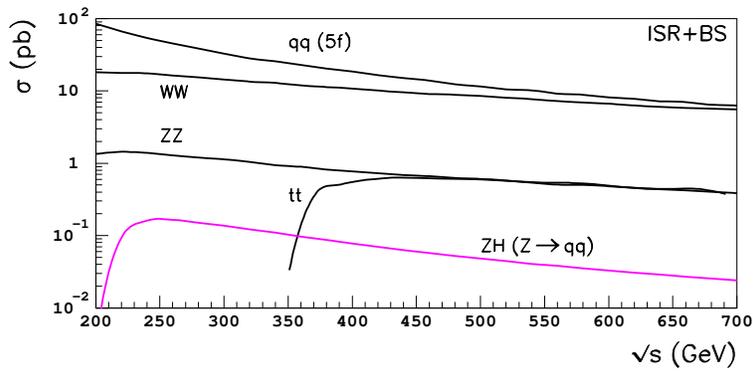,width=11cm}
}
\end{center}
\caption{\protect\footnotesize
Total cross-section for $ZH \to q\bar{q}H$ at lowest order, for $M_H = 120$ GeV, and
the main background processes considered, as a function of the center-of-mass energy.
Initial state radiation and beamstrahlung have been included.
}
\label{sigma2}
\end{figure}
%%%%%%%%%%%%%%%%%%%%%%%%%%%%%%%%%%%%%%%%%%%%%%%%%%%%%%%%%%%%%%%%%%%%%%%%%%%%
\newpage
%%%%%%%%%%%%%%%%%%%%%%%%%%%%%%%%%%%%%%%%%%%%%%%%%%%%%%%%%%%%%%%%%%%%%%%%%
%%%%%%%%%%%%%%%%%%%%%%%%%%%%%%%%%%%%%%%%%%%%%%%%%%%%%%%%%%%%%%%%%%%%%%%%%%%%
\newpage
%%%%%%%%%%%%%%%%%%%%%%%%%%%%%%%%%%%%%%%%%%%%%%%%%%%%%%%%%%%%%%%%%%%%%%%%%%%%
%%%%%%%%%%%%%%%%%%%%%%%%%%%%%%%%%%%%%%%%%%%%%%%%%%%%%%%%%%%%%%%%%%%%%%%%%%%%
\newpage
%%%%%%%%%%%%%%%%%%%%%%%%%%%%%%%%%%%%%%%%%%%%%%%%%%%%%%%%%%%%%%%%%%%%%%%%%%%%
%%%%%%%%%%%%%%%%%%%%%%%%%%%%%%%%%%%%%%%%%%%%%%%%%%%%%%%%%%%%%%%%%%%%%%%%%%%%
\newpage
%%%%%%%%%%%%%%%%%%%%%%%%%%%%%%%%%%%%%%%%%%%%%%%%%%%%%%%%%%%%%%%%%%%%%%%%%%%%
%%%%%%%%%%%%%%%%%%%%%%%%%%%%%%%%%%%%%%%%%%%%%%%%%%%%%%%%%%%%%%%%%%%%%%%%%%%%
\newpage
%%%%%%%%%%%%%%%%%%%%%%%%%%%%%%%%%%%%%%%%%%%%%%%%%%%%%%%%%%%%%%%%%%%%%%%%%%%%
%%%%%%%%%%%%%%%%%%%%%%%%%%%%%%%%%%%%%%%%%%%%%%%%%%%%%%%%%%%%%%%%%%%%%%%%%%%%
\newpage
%%%%%%%%%%%%%%%%%%%%%%%%%%%%%%%%%%%%%%%%%%%%%%%%%%%%%%%%%%%%%%%%%%%%%%%%%%%%%
%%%%%%%%%%%%%%%%%%%%%%%%%%%%%%%%%%%%%%%%%%%%%%%%%%%%%%%%%%%%%%%%%%%%%%%%%%%%
\newpage
%%%%%%%%%%%%%%%%%%%%%%%%%%%%%%%%%%%%%%%%%%%%%%%%%%%%%%%%%%%%%%%%%%%%%%%%%%%%%
%%%%%%%%%%%%%%%%%%%%%%%%%%%%%%%%%%%%%%%%%%%%%%%%%%%%%%%%%%%%%%%%%%%%%%%%%%%%
\newpage
%%%%%%%%%%%%%%%%%%%%%%%%%%%%%%%%%%%%%%%%%%%%%%%%%%%%%%%%%%%%%%%%%%%%%%%%%%%%%
%%%%%%%%%%%%%%%%%%%%%%%%%%%%%%%%%%%%%%%%%%%%%%%%%%%%%%%%%%%%%%%%%%%%%%%%%%%%
%%%%%%%%%%%%%%%%%%%%%%%%%%%%%%%%%%%%%%%%%%%%%%%%%%%%%%%%%%%%%%%%%%%%%%%%%%%%
%%%%%%%%%%%%%%%%%%%%%%%%%%%%%%%%%%%%%%%%%%%%%%%%%%%%%%%%%%%%%%%%%%%%%%%%%%%%
\newpage
%%%%%%%%%%%%%%%%%%%%%%%%%%%%%%%%%%%%%%%%%%%%%%%%%%%%%%%%%%%%%%%%%%%%%%%%%%%%%
%%%%%%%%%%%%%%%%%%%%%%%%%%%%%%%%%%%%%%%%%%%%%%%%%%%%%%%%%%%%%%%%%%%%%%%%%%%%
\newpage
%%%%%%%%%%%%%%%%%%%%%%%%%%%%%%%%%%%%%%%%%%%%%%%%%%%%%%%%%%%%%%%%%%%%%%%%%%%%%
%%%%%%%%%%%%%%%%%%%%%%%%%%%%%%%%%%%%%%%%%%%%%%%%%%%%%%%%%%%%%%%%%%%%%%%%%%%%
\newpage
%%%%%%%%%%%%%%%%%%%%%%%%%%%%%%%%%%%%%%%%%%%%%%%%%%%%%%%%%%%%%%%%%%%%%%%%%%%%%
%%%%%%%%%%%%%%%%%%%%%%%%%%%%%%%%%%%%%%%%%%%%%%%%%%%%%%%%%%%%%%%%%%%%%%%%%%%%
\newpage 
%%%%%%%%%%%%%%%%%%%%%%%%%%%%%%%%%%%%%%%%%%%%%%%%%%%%%%%%%%%%%%%%%%%%%%%%%%%%%
%%%%%%%%%%%%%%%%%%%%%%%%%%%%%%%%%%%%%%%%%%%%%%%%%%%%%%%%%%%%%%%%%%%%%%%%%%%%
\newpage 
%%%%%%%%%%%%%%%%%%%%%%%%%%%%%%%%%%%%%%%%%%%%%%%%%%%%%%%%%%%%%%%%%%%%%%%%%%%%%
%%%%%%%%%%%%%%%%%%%%%%%%%%%%%%%%%%%%%%%%%%%%%%%%%%%%%%%%%%%%%%%%%%%%%%%%%%%%
%%%%%%%%%%%%%%%%%%%%%%%%%%%%%%%%%%%%%%%%%%%%%%%%%%%%%%%%%%%%%%%%%%%%%%%%%%%%%
%%%%%%%%%%%%%%%%%%%%%%%%%%%%%%%%%%%%%%%%%%%%%%%%%%%%%%%%%%%%%%%%%%%%%%%%%%%%
\newpage
%%%%%%%%%%%%%%%%%%%%%%%%%%%%%%%%%%%%%%%%%%%%%%%%%%%%%%%%%%%%%%%%%%%%%%%%%%%%%
%%%%%%%%%%%%%%%%%%%%%%%%%%%%%%%%%%%%%%%%%%%%%%%%%%%%%%%%%%%%%%%%%%%%%%%%%%%%


\newpage

\begin{thebibliography}{99}

\newcommand{\mpl}{Mod. Phys. Lett.}
\newcommand{\cpc}{Comp. Phys. Com.}
\newcommand{\np}{Nucl. Phys.}
\newcommand{\prl}{Phys. Rev. Lett.}
\newcommand{\pr}{Phys. Rev.}
\newcommand{\nim}{Nucl. Inst. and Meth.}
\newcommand{\pl}{Phys. Lett.}
\newcommand{\zp}{Z. Phys.}
\newcommand{\y}{{\sl et al.}}

\bibitem{SM} S.L. Glashow, {\np} {\bf B22} (1961) 579; \\
             S. Weinberg, {\prl} {\bf 19} (1967) 1264; \\
             A. Salam, {\em Elementary Particle Theory}, ed. N. Svartholm, 
             Almqvist and Wiksell, Stockholm (1968) 367; \\
	     S.L. Glashow, J. Iliopoulos and L. Maiani, {\pr} {\bf D2} (1970) 1285.
\bibitem{Peskin} H. Murayama and M.E. Peskin, hep-ex/9606003, June 10, (1996).
\bibitem{higgstrahlung} J. Ellis, M.K. Gaillard, D.V. Nanopoulos, {\np} {\bf B106} (1976) 292; \\
                        J.D. Bjorken, Proc. SLAC Summer Institute, 1976; \\ 
                        B. Ioffe, V. Khoze, Sov. J. Part. Nucl. Phys. {\bf 9} (1978) 50; \\
                        B.W. Lee, C. Quigg, H.B. Thatcher, {\pr} {\bf D16} (1977) 1519.
\bibitem{fusion} D.R.T. Jones, S.T. Petcov, {\pl} {\bf 84B} (1979) 440; \\
                 R.N. Cahn, S. Dawson, {\pl} {\bf 136B} (1984) 196; \\
                 G. Altarelli, B. Mele, F. Pitolli, {\np} {\bf B287} (1987) 205. 
\bibitem{lepewwg} The LEP Collaborations, CERN-EP-99-015.
\bibitem{limit}   ALEPH 99-081 CONF 99-052; 
                  DELPHI 99-142 CONF 327; \\
                  L3 Note 2442; 
                  OPAL Technical Note TN-614.
\bibitem{djouadi} A. Djouadi, hep-ph/99010449, October 22, (1999).
\bibitem{previous} P. Janot in
                   {\em Physics and Experiments with Linear $e^+e^-$ Colliders},
                   Harris F.A., Olsen S.L., Pakvasa S. and Tata X., eds. (World
		   Scientific, Singapore, 1993).
\bibitem{jetset} T. Sj\"{o}strand, {\cpc} {\bf 82} (1994) 74.
\bibitem{circe} T. Ohl, hep-ph/9607454-rev, July 1996 (expanded September 1996).
\bibitem{simdet} M. Pohl and H.J. Schreiber, DESY Preprint DESY 99-01, January 1999.
\bibitem{nlccdr} Webpage for the Model Detectors of the USA/Canada Linear
	         Collider Detector Simulation Group:
	         ``http://www.slac.stanford.edu/$^{\sim}$mpeskin/LC/modeld.html''.
\bibitem{bramhs} T. Behnke, G. Blair, K. M\"{o}ning and M. Pohl, November 8 , (1998), \\
                 ``http://www.hep.ph.rhbnc.ac.uk/$^{\sim}$blair/detsim/brahms.html''.
\bibitem{aleph} D. Brown and M. Frank, ALEPH 92-135 PHYSIC 92-124.
\bibitem{pablo} P. Garc\'\i a-Abia and W. Lohmann, hep-ex/9908065, August 30, (1999).
\bibitem{optproj} Ll. Garrido, S. G\'omez, A. Juste and V. Gaitan, {\cpc} {\bf 104} (1997) 37.
\bibitem{JADE} W. Bartel {\y} (JADE Collaboration), {\zp} {\bf C33} (1986) 23.
\bibitem{LUCLUS} T. Sj\"{o}strand, {\cpc} {\bf 22} (1983) 227.
  
\end{thebibliography}
\end{document}